\documentclass{LT23auth}
\usepackage{graphicx}
\def\Vec#1{\mbox{\boldmath $#1$}}

\begin{document}

\begin{frontmatter}

\title{Competition between rotation and turbulence in superfluid He$^4$}

\author[address1]{Tsunehiko Araki\thanksref{thank1}},
\author[address1]{Makoto Tsubota},
\author[address2]{Carlo F. Barenghi}

\address[address1]{Department of Physics, Osaka City University, Sumiyoshi-Ku, Osaka 558-8585, Japan}

\address[address2]{Department of Mathematics, University of Newcastle, Newcastle upon Tyne, NE1 7RU, UK}

\thanks[thank1]{ E-mail:taraki@sci.osaka-cu.ac.jp}

\begin{abstract}
Two types of vortex states have been much studied in superfluid
$^4$He. The first is the vortex array in a rotating container. The second is the
vortex tangle in turbulent flow. An experiment attempt to combine these two states
by rotating a counterflow was attempted years ago. The data suggest the existence of
different flow regimes separated by instabilities, but a theoretical interpretation
is still missing. We present work in which we use the vortex filament model to
numerically investigate rotating counterflow. We show evidence of a new state of
polarized turbulence.
\end{abstract}

%
%
\end{frontmatter}

\section{Introduction}

The first studies of quantized vorticity involved rotating a sample of $^4$He at constant angular velocity, and resulted in an ordered array of vortices aligned along the rotation axis \cite{Donnelly1}. Quantized vortices can also be created by applying a thermal counterflow \cite{Tough1}, in which the vortices form an irregular, disordered tangle of lines. There has been only one experiment that we are aware of in which these two methods of producing vortices are combined \cite{Swanson1}. The experiment suggests that there exists a form of steady rotating turbulence, characterized by a certain vortex line density at given counterflow velocity $v_{\rm ns}$ and angular velocity $\Omega$.

The aim of this work is to determine numerically in the first place whether such a state of rotating turbulence exists.

\section{Numerical calculation}

We assume the model of Schwarz \cite{Schwarz1} in which a quantized vortex line is described as a parametric form $\Vec{s}=\Vec{s}(\xi,t)$. Then the velocity of a point $\Vec{s}$ is given by
\begin{equation}
\dot{\Vec{s}}=\dot{\Vec{s}}_0 + \alpha \Vec{s}' \times (\Vec{v}_{\rm ns} -\dot{\Vec{s}}_0), \label{eq.1}
\end{equation}
where $\dot{\Vec{s}}_0$ is determined by the fully non-local Biot Savart law, and $\alpha$ is the temperature-dependent friction coefficient. The numerical simulation is performed in a cubic box of volume $L^3=1$ cm$^3$ in a rotating frame at angular velocity $\Vec{\Omega}=\Omega \hat{ \Vec{z}}$ at temperature $T=$1.6 K. We assume periodic boundary conditions along the $z$ direction and rigid boundary conditions at the side walls.

\section{Results}

We calculate the dynamics of vortices at $\Omega=6.3471 \times 10^{-3}$ Hz and $v_{\rm ns}=0.08$ cm/s (Fig. 1) \cite{movie}. We create an initial vortex array (Fig. 1a) which consists of 8 vortices parallel to the rotation axis. We added very small random perturbations to the initial configuration of vortices (of rms amplitude 0.01 cm) to simulate thermal or mechanical noise. The perturbations undergo the Glaberson instability \cite{Glaberson1} and Kelvin waves (helical displacements of the vortex cores) become unstable and grow, as shown in Fig. 1b. The randomness of the initial perturbations prevents the locking of the phases of these growing  waves. Figure 2 shows that the vortex line density $L$ grows exponentially, as predicted by Glaberson {\it et al.}, starting from its initial value $L(0) \simeq 8$ cm$^{-2}$. Soon the helical waves reconnect with each other (Fig. 1c) and a vortex tangle is formed (see Fig. 1d), reaching a saturated value of the vortex line density $L\simeq 40$ cm$^{-2}$ as seen in Fig. 2.

\begin{figure}[tbhp]
\begin{minipage}{1.0\linewidth}
\begin{center}
\includegraphics[width=0.8\linewidth]{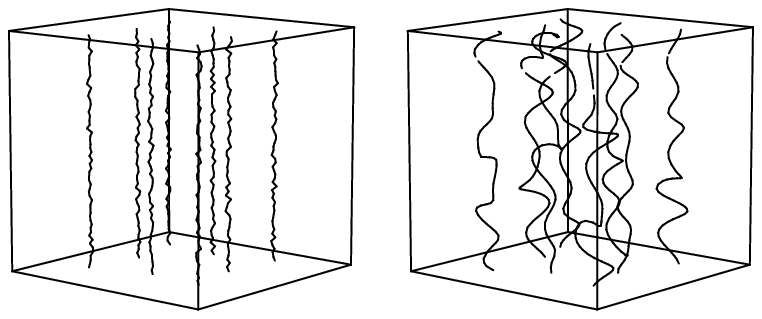}\\
 (a) \hspace{2cm} (b)\\
\end{center}
\end{minipage}

\begin{minipage}{1.0\linewidth}
\begin{center}
\includegraphics[width=0.8\linewidth]{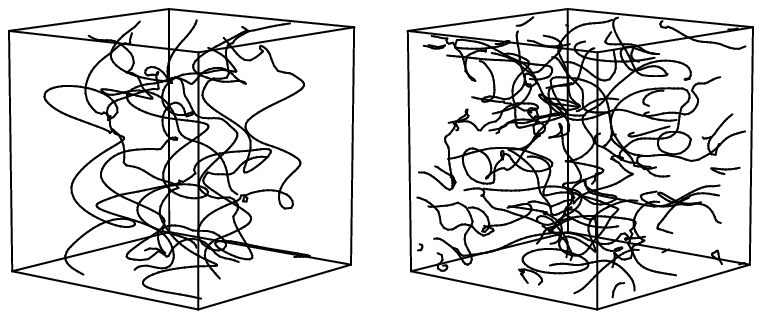}\\
 (c) \hspace{2cm} (d)\\
\end{center}
\end{minipage}
\caption{Time evolution of vortices at $t$=0 sec(a), $t$=24.0 sec(b),
$t$=40.0 sec(c) and $t$=200.0 sec(d).}
 \label{eps1}
\end{figure}

\begin{figure}[tbhp]
\begin{center}
\includegraphics[width=0.8\linewidth]{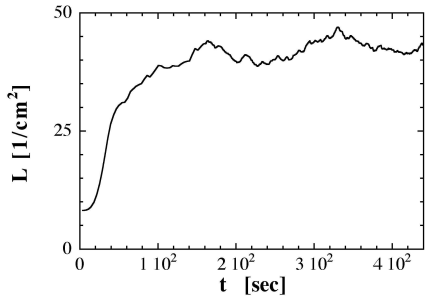}\\
\end{center}
\caption{Time evolution of the vortex line density.}
 \label{eps2}
\end{figure}

To analyze the results we compute the average projection of the tangent unit vector along the vortex lines in the rotation direction, $<\Vec{s}'_z>$, shown in Fig. 3. At $t=0$ $<\Vec{s}'_z>=1$ because all lines are in the rotation direction. As the vortex array is randomized by the Glaberson instability and the resulting reconnections, $<\Vec{s}'_z>=1$ decreases from one as the tangle builds up in intensity. It is however apparent that when a steady state is reached the vortex tangle, although apparently random (see Fig. 1d), possesses a definite polarization $<\Vec{s}'_z> \simeq 0.15$. This result supports the interpretation of the experiment given by Swanson {\it et al.}, that superfluid turbulence can be polarized.

\begin{figure}[tbhp]
\begin{center}
\includegraphics[width=0.8\linewidth]{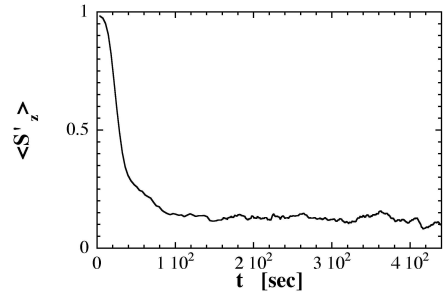}\\
\end{center}
\caption{Time evolution of the average projection of the tangent unit vector along the vortex lines in the rotation direction.}
 \label{eps3}
\end{figure}

\section{Conclusion}

We have studied numerically the vortex dynamics under both rotation and counterflow. We find that the initial rotating array is made unstable by a counterflow velocity as predicted by Glaberson {\it et al.}, and that, after many reconnections, a vortex tangle is formed. The tangle, apparently random, actually possesses a definite polarization. This is the first numerical evidence for the existence of a polarized state of superfluid turbulence.

%
%
  
%
%
\begin{ack}
We acknowledge W.F. Vinen and D.C. Samuels for useful discussions. This study is supported by the Japan - UK Scientific Cooperative
Program (Joint  Research Project) by JSPS and Royal Society.
\end{ack}

%
%

\end{document}